\documentstyle[aps,epsf,prb,twocolumn,floats]{revtex}

\begin{document}

\title{
Highly Degenerate Canted Spin Structure\\
in Bilayer Manganite La$_{1.1}$Sr$_{1.9}$Mn$_2$O$_7$
}

\author{
M. Kubota,~\cite{kubota} Y. Oohara, and H. Yoshizawa}
\address{Neutron Scattering Laboratory, I. S. S. P., University of
Tokyo, Tokai, Ibaraki, 319-1106, Japan
 }
\author{
H. Fujioka and K. Hirota
}
\address{
CREST, Department of Physics, Tohoku University, Aoba-ku, Sendai,
980-8578, Japan
}
\author{
Y. Moritomo
}
\address{PRESTO and CIRSE, Nagoya University, Nagoya, 464-8601, Japan
}
\author{
Y. Endoh
}
\address{Institute for Materials Research, Tohoku University, Aoba-ku,
Sendai
980-8577, Japan 
}

\date{\today}

\wideabs{

\maketitle
\begin{abstract}
We have examined magnetic ordering of the two-dimensional (2D) bilayer
manganite La$_{1.1}$Sr$_{1.9}$Mn$_2$O$_7$  with neutron diffraction
technique.  Due to the two-dimensionality, there appears intense 2D ridge
scattering along the tetragonal $(00l)$ direction, and the analyses of such
scattering leads to a conclusion that the low temperature spin structure
is a canted antiferromagnetic ordering.
\end{abstract}

\pacs{75.25.+z, 75.30.Vn, 71.70.Ej}
}




It is well known that an appropriate hole doping in a manganite, {\it e.g.}
La$_{1-x}$Ca$_{x}$MnO$_{3}$, turns an antiferromagnetic (AFM) insulator
to a ferromagnetic (FM) metal.  In a very early study by Wollan and
Koehler, on the other hand, it was already pointed out that an AFM ordering
survives in the metallic FM phase.~\cite{wollan}  de Gennes successfully
demonstrated that the coexistence of the ferromagnetism and the
antiferromagnetism can be interpreted as a formation of a canted spin
structure through the competition between the superexchange AFM
interaction and the FM double-exchange interaction.~\cite{P.G.deGennes60} 
Based on the NMR study, however, Allodi {\it et al.} recently claimed that
the coexistence should be interpreted as a phase separation of the FM
metal phase and the AFM insulator phase.~\cite{Allodi}   Theoretical
studies by Moreo {\it et al.} suggest that the phase separation is favored
against the canted spin structure as a ground state of the hole-doped
manganite.~\cite{moreo}  Since then, a number of researches were carried
out to explore whether the phase separation really exists in a hole-doped
manganite system.


La$_{2-2x}$Sr$_{1+2x}$Mn$_2$O$_7$ is known as one of colossal
magnetoresistance (CMR) manganites.  In particular, the $x=0.40$ sample
exhibits a dramatic decrease of the resistivity by applying a magnetic
field just above $T_{\rm C}$, and the phase separation picture is invoked
as an origin of such behavior.  La$_{2-2x}$Sr$_{1+2x}$Mn$_2$O$_7$  is a
two-dimensional (2D) bilayered manganite with a tetragonal body-
centered structure.\cite{Moritomo96,Mitchell97}  Between $x=0.4$ and
$x=0.48$, it exhibits two magnetically ordered phases; the AFM insulator
for $T_{\rm C} \leq T \leq T_{\rm N}$, and the FM metal  below $T_{\rm
C}$.~\cite{Kubota99a,Kubota99b} Based on the neutron scattering
experiments, the low $T$ spin structure is reported to be a canted AFM
ordering.  Due to the 2D layered structure, an overall spin structure
consists of units of ferromagnetically ordered $c$-planes; the two FM
layers couple ferromagnetically within a bilayer, while such FM bilayer
units couple antiferromagnetically.~\cite{Hirota98,Kubota00b}  What was
actually observed in our previous neutron diffraction studies is the
coexistence of the AFM as well as FM Bragg intensities.  Conceptually,
however, such observations allow a possibility of the phase separation
because it is also compatible with diffraction data.  In fact, Moreo {\it et
al.} argued that such a coexistence of FM and AFM Bragg reflections should
be considered as evidence of the phase separation and suggested that the
existence of the FM metallic clusters in the A-type AFM phase is the
origin of the CMR in this compound.~\cite{Moreo99}  On the other hand,
Maezono and Nagaosa suggested that the two-dimensionality of the
La$_{2-2x}$Sr$_{1+2x}$Mn$_2$O$_7$ system allows a possibility of a
canted spin structure.~\cite{maezono}


In the present work, we shall demonstrate that, by taking an advantage of
the 2D nature of the La$_{2-2x}$Sr$_{1+2x}$Mn$_2$O$_7$ system, it is
possible to distinguish the canted AFM ordering from the phase separation
by a diffraction study.  We have observed magnetic diffuse signals in the
$x=0.45$ sample in the paramagnetic (PM) phase, an A-type AFM phase and
the low $T$ phase by neutron diffraction technique.  From the analyses of
the diffuse scattering, more precisely the 2D ridge scattering, we
demonstrate that the low $T$ spin structure is, indeed, a canted spin
structure.


The sample studied is the same with that used in our series of the
previous studies.~\cite{Hirota98,Kubota00b,Kubota00a}  Neutron
scattering measurements were carried out with the ISSP triple-axis
spectrometer GPTAS installed at the 4G experimental port of JRR-3M in
JAERI (Tokai).  We used neutrons with $k_{i}=3.826$~\AA$^{-1}$ and
employed the collimation of 40'-40'-40' in the double-axis mode.  A
monochromatic beam was obtained by the $(002)$ reflection from
pyrolytic graphite.  To remove higher order neutrons, we used a pyrolytic
graphite filter.  The sample was placed in an Al can with the $[010]$ axis
vertical to observe the magnetic signal in the $(h~0~l)$ scattering plane,
and it was cooled with a closed cycle helium gas refrigerator down to 10
K.



\begin{figure}[htb]
\begin{center}
\leavevmode
\epsfxsize=0.85\linewidth
\epsfbox{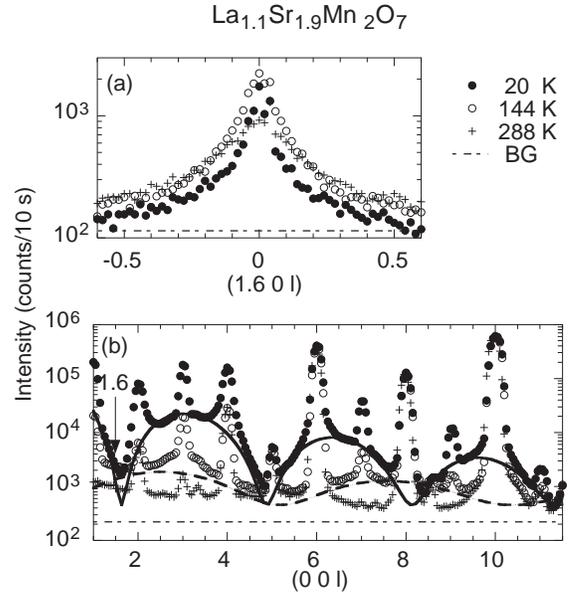}
\end{center}
\caption{(a) $(0~0~l)$ scan at 288~K, 144~K and 20~K. The grey line is
proportional
to $-\cos(2 \pi  z_{1}l)+const.$ and the solid line to $\cos(2 \pi 
z_{2}l)+const.$, where they are corrected by the Lorentz factor. (b)
(h~0~1.6)
scan at 288~K, 144~K and 20~K.}
\label{scan}
\end{figure}

Let us start from the FM spin correlations within MnO$_2$ layers (the
$ab$ plane).  Figure \ref{scan}(a) is the profiles measured along the
$(100)$ direction.  The peak observed at $h =0$ indicated that, reflecting
the two-dimensionality of the system, there is a well-developed FM spin
correlations in each MnO$_2$ layer even in the PM phase at 288 K, and it
further develops with lowering $T$.  In Fig. \ref{scan}(b) are shown the
profiles which gives information on the spin correlations perpendicular to
the MnO$_2$ bilayers measured along the $(0~0~l)$ line at 288~K in the
PM phase, at 144~K in the intermediate A-type AFM phase, and at 20~K in
the low $T$ canted AFM phase, respectively.  Due to the very good two-
dimensionality of the spin system, there is practically no correlation
between bilayers in the PM phase.  The AFM Bragg reflections are observed
at $Q = (0~0~2n+1)$ at 144~K in the intermediate A-type AFM phase,
while at 20K in the canted AFM phase, FM as well as AFM Bragg reflections
are observed at $Q = (0~0~2n)$ and $(0~0~2n+1)$, respectively.  In
addition, distinct sinusoidal modulations are observed along the $(0~0~l)$
line in the intermediate A-type AFM as well as low $T$ canted AFM
phases, reflecting the bilayer nature of the La$_{2-
2x}$Sr$_{1+2x}$Mn$_2$O$_7$ system.  It should be noted, however, that
the periodicity of the modulation in two phases is quite different.

Through the careful examination of the profiles depicted in Fig. 1(b), one
can easily find that the sinusoidal modulation at 144~K in the AFM phase
is proportional to $-\cos(2 \pi  z_{1}l)$, where $z_{1}c$ is the distance
between the MnO$_2$ intra-bilayers. This means that within each
bilayers, the MnO$_2$ layers are  {\it antiferromagnetically} coupled as
depicted in Fig. 2(a).  By contrast, the sinusoidal modulation at 20 K in the
canted AFM phase is proportional to $\cos(2 \pi  z_{2}l)$, where $z_{2}c$
is the distance between the {\it inter-bilayers} depicted in Fig. 2(b).  This
behavior is unexpected because one expects that the two-dimensionality
of this system favors the strong AFM coupling between MnO$_2$ layers
within a bilayer-unit, while loose correlation between
antiferromagnetically coupled bilayer-units.  The $\cos(2 \pi  z_{2}l)$
modulation undoubtedly indicates that the `AFM block'  in Fig. 2(a) is not
effective, and instead the `FM coupling' between the bilayer-units shown
in Fig. 2(b) controls the spin correlation perpendicular to the MnO$_2$
layers in the low $T$ canted AFM phase.

The well-developed FM spin correlation throughout all $T$ (Fig. 1(a))
makes the analysis of the spin correlations along the $(0~0~l)$ line
considerably simple.  It allows us to define a thermal average of the spin
direction within each layer, and the spins in the $j$-th layer may be
expressed as,
\begin{equation}
S_{j}^{+}=\langle S\rangle \exp\imath \phi_{j},
\end {equation}
where $S_{j}^{+}=S_{j}^{x}+i S_{j}^{y}$, and $\phi_{j}$ stands for the spin
direction of the spins in the $j$-th layer relative to the $x$ axis.  With
these assumptions, the analysis of the spin structure is reduced to a one-
dimensional problem, and the scattering profile along the $(0~0~l)$ line
may be expressed by
\begin{eqnarray}
\frac{\rm d \sigma}{\rm d \Omega}&=&N_{c}r_{0}^{2}
\left (\frac{1}{2}g\langle S
\rangle f(k)\right )^{2}\left (\frac{2\pi}{a}\right )^{2}\nonumber\\
\times&\sum_{i,j}&\langle \exp \left (\imath (\phi_{i}-\phi_{j})
\right)\rangle
\exp\left (2\pi \imath  l z_{ij}\right )\delta(h)\delta(k)+\cdots.
\end{eqnarray}
where parameters have their usual meaning, \cite{lovesey} $N_{c}$ is the
number of Mn site in an  FM $c$-plane and $z_{ij}c$ is the distance
between
the $i$-th and $j$-th FM $c$-plane.  The term $\langle \exp \left (\imath
(\phi_{i}-\phi_{j})\right)\rangle$ denotes an angle correlation function.

It is straightfoward to derive an effective scattering cross-section which
describes the $l$ dependence of the magnetic scattering observed along
the 2D ridge on the $(0~0~l)$ line shown in Fig. 1(b).  Suppose that the
antiferromagnetic interaction $J_{1}$ forms the AFM block depicted in
Fig. 2(a),  we can evaluate the angle correlation between the $n+k$-- and
the $n$--th  AFM block as,

\begin{eqnarray}
\label{correlation function}
\langle \exp \left (\imath (\phi_{n+k}-\phi_{n})
\right)\rangle=(-r)^{k}.
\end{eqnarray}
Here, $\phi_{n}$ stands for the spin direction of the lower layer of the
$n$-- th AFM block and $r$ is the coherence parameter; $r=1$ for the long
range ordering while $r=0$ when there is no correlation between the
bilayers.  By using this correlation function, the sinusoidally modulated
diffuse scattering in the intermediate AFM phase can be expressed by,
\begin{eqnarray}
\label{cs1}
\frac{\rm d \sigma}{\rm d \Omega}&=&2Nr_{0}^{2}\left
(\frac{1}{2}g\langle S \rangle f(k)\right )^{2}\left
(\frac{2\pi}{a}\right)^{2} \{1-\cos(2\pi l z_{1})\}  \nonumber\\
 &\times& \frac{1-r^{2}}{1+2r\cos(\pi l)+r^{2}}\delta(h)\delta(k)+\cdots,
\end{eqnarray}
where $N$ is the number of Mn site in the sample.
For the perfect correlation between bilayer blocks, we note that this
formula is reduced to the following equation, which describes the AFM
Bragg reflections observed at $(0~0~2n+1)$ in the AFM phase (Fig. 1(a)),

\begin{eqnarray}
\frac{\rm d \sigma}{\rm d \Omega}&=&4Nr_{0}^{2}
\left (\frac{1}{2}g\langle S \rangle  f(k)\right )^{2} 
\frac{(2\pi)^{3}}{a^{2}c}\left \{2\sin(\pi l z_{1})\right \}^{2}  \nonumber\\
 &\times& \sum_{n} \delta\left (l-(2n+1)\right)\delta(h)\delta(k)+\cdots.
\end{eqnarray}

Next, we discuss the cross-section for the unusual modulation observed in
the low $T$ canted AFM phase.  Suppose that the interaction $J_{1}'$
makes a
canting angle $\theta$ between the adjacent spins on two layers within a
bilayer and the ferromagnetic interaction $J_{2}$ forms FM blocks  as
shown in Fig. 2(b) .  Then, the spin canting angle between the $n$--th and
$(n+1)$--th FM blocks is expressed as,

\begin{equation}
\label{angle}
\phi_{n+1}-\phi_{n}=\pm \theta.
\end{equation}

Note that the spin direction can turn either $+\theta$ or $-\theta$ when
going to adjacent bilayer FM blocks. Namely, the ground state generated by
$J_{1}'$ and $J_{2}$ includes a high degeneracy of the turn angle of the
spin direction. The long range order of  a canted  spin-structure can be
established by introducing the ferromagnetic-interaction of the 
third-nearest exchange interaction $J_{3}$ depicted in Fig. 2(b), which
lifts the
degeneracy of the ground state as discussed by de
Gennes.~\cite{P.G.deGennes60}

\begin{figure}[htbp]
\begin{center}
\leavevmode
\epsfxsize=80mm
\epsfxsize=0.85\linewidth
\epsfbox{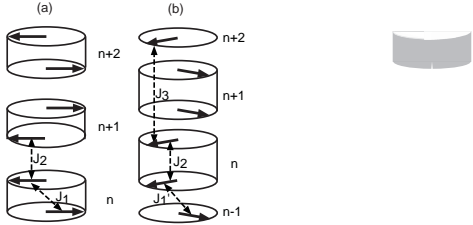}
\end{center}
\caption{
(a): Antiferromagnetic coupling within bilayer block $J_{1}$ and $J_{2}$
denotes the intra-bilayer and inter-bilayer exchange interactions.
(b):  Spin canting within bilayer block units by $J_{1}'$ and ferromagnetic
coupling between bilayer blocks by $J_{2}$. $J_{3}$ denotes the exchange
interaction between every other bilayer-units. }
\label{cant}
\end{figure}

For this one-dimensional canted spin structure, we can calculate the angle
correlation function $\left \langle\exp(\imath(\phi_{n+k}-
\phi_{n}))\right \rangle$ from the analogy of the calculation for the 
magnetization in the one-dimensional Ising spin system as,~\cite{mattis}
\begin{eqnarray} 
\left \langle \exp(\imath(\phi_{n+k}-
\phi_{n}))\right \rangle&=&\left \langle  \exp \left (\imath
\sum_{j=n}^{k-1}\phi_{j+1}-\phi_{j}\right
)\right \rangle\nonumber\\
&=&A_{+}r_{+}^{k}+A_{-}r_{-}^{k},
\end{eqnarray}
where $A_{+}, r_{+}, A_{-}$ and $r_{-}$ are given by the canting angle
$\theta$ and the coherence parameter $p$ as,

\begin{eqnarray}
A_{\pm}=\frac{1}{2}\pm\frac{(1-
p)\cos\theta}{2\sqrt{(p\cos\theta)^{2}-2p+1}},\\
r_{\pm}=p\cos\theta \pm\sqrt{(p\cos\theta)^{2}-2p+1}.
\end{eqnarray}
Here, the coherence parameter $p$
denotes the coherence between the FM blocks; $p=0$ for the long-range
helical order while $p= \frac{1}{2}$ for a random sequence of the canting
angles (the degenerate ground state).

By using these formula for the correlation function, we obtain the
following cross-section for the diffuse scattering in the low $T$ canted
AFM structure,
\begin{eqnarray}
\label{cs2}
\frac{\rm d \sigma}{\rm d \Omega}&=&2Nr_{0}^{2}
\left (\frac{1}{2}g\langle S
\rangle f(k)\right )^{2}\left (\frac{2\pi}{a}\right )^{2}
\left \{2\cos(\pi l z_{2})\right\}^{2}\nonumber\\
 &\times &\left [\frac{A_{+}(1-r_{+}^{2})}
{1-2r_{+}\cos(\pi l)+r_{+}^{2}}
+\frac{A_{-}(1-r_{-}^{2})}
{1-2r_{-}\cos(\pi l)+r_{-}^{2}}\right ]\nonumber\\
&\times  &\delta(h)\delta(k)+\cdots.
\end{eqnarray}

When the FM correlation between the next-nearest blocks is well
developed, this equation is reduced to the following cross-section which
describes the AFM and FM Bragg reflections observed in the low $T$ canted
AFM phase,

\begin{eqnarray}
\frac{\rm d \sigma}{\rm d \Omega}&=&4Nr_{0}^{2}\left
(\frac{1}{2}g\langle S\rangle f(k)\right )^{2}\frac{(2\pi)^{3}}{a^{2}c}
\left \{2\cos(\pi l z_{2})\right \}^{2}\nonumber\\
&\times&\sum_{n}\left [
\left (\cos\frac{\theta}{2}\right )^{2}\delta(l
-2n)+\left (\sin \frac{\theta}{2
}\right )^{2}\delta\left(l-(2n+1)\right)\right]\nonumber\\
&\times&\delta(h)\delta(k)+\cdots.
\end{eqnarray}

\begin{figure}[htbp]
\begin{center}
\leavevmode
\epsfxsize=0.85\linewidth
\epsfbox{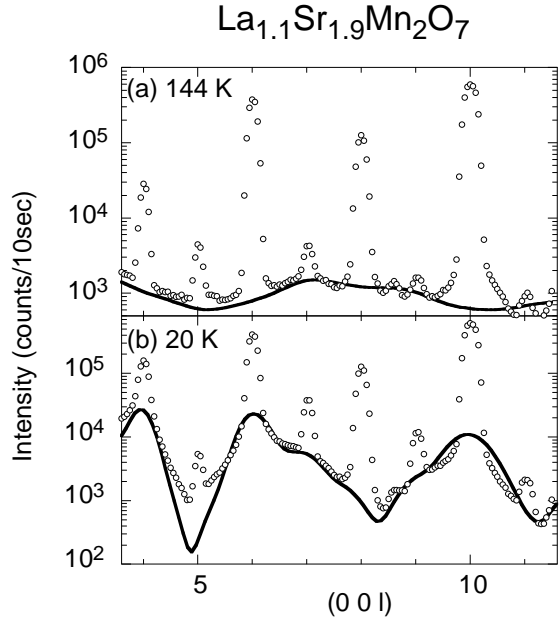}
\end{center}
\caption{
Observed intensity vs. calculated intensity at (a) 144 K and (b) 20 K.
}
\label{fit}
\end{figure}

With using the structural parameters $\theta= 63.7^{\circ}$,
$z_{1}=0.194$ and $z_2=0.306$ which are obtained in our previous
studies,\cite{Hirota98,Kubota00b} and with the magnetic form factor of
Mn$^{3+}$,~\cite{brown} the $l$ dependences of the diffuse scattering
observed in the intermediate AFM phase and the low $T$ canted AFM phase
shown in Fig. 1(b) were fitted to the equations derived above with the
instrumental function convoluted.   Since most of the structural
parameters are known, only one parameter is left to be determined: the
coherence parameter $r$ in Eq. (\ref{cs1}) for the intermediate AFM phase,
and $p$ in Eq.(\ref{cs2}) for the low $T$ canted AFM phase, respectively. 
Figure ~\ref{fit} shows the observed data and the calculated intensity
with parameters, $r=0.10$ and $p=0.38$, respectively.  It is gratifying
that the sinusoidal-shape with the precise periodicity of a diffuse signal
is well reproduced by these models for both phases.  An important
consequence of the present analyses is that the sinusoidal modulation of
diffuse scattering in the intermediate A-type AFM state is characterized
by the coupling of the ferromagnetically ordered bilayers within bilayer-
units shown in Fig. 2(b), whereas that in the low $T$ canted AFM state is
consistently described by taking account of the high degeneracy of the
canted spin structure.  We conclude that the $\cos(2 \pi  z_{2}l)$
modulation in the low $T$ phase provides strong evidence of the existence
of the canted spin structure in the low $T$ phase.

In conclusion, we have observed the diffuse magnetic signals in
La$_{1.1}$Sr$_{1.9}$Mn$_2$O$_7$ for three representative phases.  The
$l$ dependences of the 2D ridge scattering observed along the $(00l)$ line
at three phases are characterized, respectively, by the two dimensionality
of the spin correlation in the PM phase, by a coupling within bilayer-units
for those in the intermediate AFM phase, and by the high degeneracy which
is inherent to the canted-spin ordering for the low $T$ canted AFM phase. 
The present results demonstrate that  the ``de Gennes" canted AFM spin
structure is formed in La$_{2-2x}$Sr$_{1+2x}$Mn$_2$O$_7$.

\end{document}